\documentclass[a4paper,twocolumn,11pt,accepted=2021-06-04]{quantumarticle}
\pdfoutput=1

\usepackage[numbers,sort&compress]{natbib}
\usepackage{graphicx, color, graphpap}% Include figure files
\usepackage{enumitem}
\usepackage{amssymb}
\usepackage{amsthm}
\usepackage{multirow}
\usepackage[colorlinks=true,citecolor=blue,linkcolor=magenta]{hyperref}
\usepackage[T1]{fontenc}
\usepackage{bbm}
\usepackage{thmtools,thm-restate}
\usepackage{verbatim}
\usepackage{mathtools}
\usepackage{titlesec}
\usepackage{amsmath}
\usepackage[title]{appendix}

\usepackage[caption = false]{subfig}

\long\def\ca#1\cb{} %Use for commenting out: \ca...\cb

\newcommand{\ket}[1]{|#1\rangle}               %ket
\newcommand{\bra}[1]{\langle #1|}              %bra
\newcommand{\dya}[1]{\ket{#1}\!\bra{#1}}
\newcommand{\poly}{\operatorname{poly}}
\newcommand{\opt}{\text{opt}}

\newcommand{\OC}{\mathcal{O}}

\newcommand{\SC}{\mathcal{S}}

\newcommand{\Tr}{{\rm Tr}}

\newcommand{\Var}{{\rm Var}}

\renewcommand{\geq}{\geqslant}
\renewcommand{\leq}{\leqslant}

\DeclareMathOperator*{\argmin}{arg\,min}
\renewcommand{\vec}[1]{\boldsymbol{#1}}  % Bold vectors instead of arrow vectors

\newcommand{\ad}{^\dagger}
\newcommand*{\id}{\openone}

\newcommand{\thv}{\vec{\theta}}

{}%         Body font
{}%         Indent amount (empty = no indent, \parindent 
\newtheorem{corollary}{Corollary}
\newtheorem{proposition}{Proposition}

\newtheorem{definition}{Definition}
\begin{document}

\title{Effect of barren plateaus on gradient-free optimization}

\author{Andrew Arrasmith} 
\thanks{The first two authors contributed equally to this work.}
\address{Theoretical Division, Los Alamos National Laboratory, Los Alamos, NM 87545, USA}

\author{M. Cerezo}
\thanks{The first two authors contributed equally to this work.}
\affiliation{Theoretical Division, Los Alamos National Laboratory, Los Alamos, NM 87545, USA}
\affiliation{Center for Nonlinear Studies, Los Alamos National Laboratory, Los Alamos, NM, USA
}

\author{Piotr Czarnik} 
\address{Theoretical Division, Los Alamos National Laboratory, Los Alamos, NM 87545, USA}

\author{Lukasz Cincio} 
\address{Theoretical Division, Los Alamos National Laboratory, Los Alamos, NM 87545, USA}

\author{Patrick J. Coles}
\affiliation{Theoretical Division, Los Alamos National Laboratory, Los Alamos, NM 87545, USA}

\begin{abstract} 
Barren plateau landscapes correspond to gradients that vanish exponentially in the number of qubits. Such landscapes have been demonstrated for variational quantum algorithms and quantum neural networks with deep circuits, global cost functions, large entanglement, or hardware noise. For obvious reasons, it is expected that gradient-based optimizers will be significantly affected by barren plateaus. However, whether or not gradient-free optimizers are impacted is a topic of debate, with some arguing that gradient-free approaches are unaffected by barren plateaus. Here we show that, indeed, gradient-free optimizers do not solve the barren plateau problem. Our main result proves that cost function differences, which are the basis for making decisions in a gradient-free optimization, are exponentially suppressed in a barren plateau. Hence, without exponential precision, gradient-free optimizers will not make progress in the optimization. We numerically confirm this by training a parameterized quantum circuit in a barren plateau landscape with several gradient-free optimizers (Nelder-Mead, Powell, and COBYLA algorithms), and show that the number of shots required in the optimization grows exponentially with the number of qubits. These results provide new insight into the training landscapes of quantum neural networks and will inform the development of strategies to mitigate or avoid barren plateaus.
\end{abstract}

\maketitle

\section{Introduction}

Parameterized quantum circuits offer a flexible paradigm for programming Noisy Intermediate Scale Quantum (NISQ) computers. These circuits are utilized in both Variational Quantum Algorithms (VQAs)~\cite{cerezo2020variationalreview,bharti2021noisy,peruzzo2014variational,mcclean2016theory,farhi2014quantum,romero2017quantum,khatri2019quantum,larose2019variational,arrasmith2019variational,cerezo2020variationalfidelity,cirstoiu2020variational,sharma2019noise,bravo2020variational,cerezo2020variational} and Quantum Neural Networks (QNNs)~\cite{schuld2014quest,cong2019quantum,beer2020training,verdon2018universal}. Both VQA and QNN approaches involve efficiently evaluating a cost function $C(\thv)$ or its gradient $\nabla C(\thv)$ on a quantum computer. A classical optimizer is then employed to train the parameters $\thv$ of a parameterized quantum circuit $V(\thv)$ to minimize the cost.

Rigorous scaling results are urgently needed for this exciting approach to near-term quantum computing. Gradient scaling is one of the few directions of significant progress. The most famous gradient scaling result is the barren plateau phenomenon~\cite{mcclean2018barren,cerezo2020cost,sharma2020trainability,wang2020noise,cerezo2020impact,holmes2020barren,pesah2020absence,zhang2020toward,abbas2020power,marrero2020entanglement,patti2020entanglement,uvarov2020barren,holmes2021connecting,du2020learnability,arrasmith2021equivalence}, whereby the gradient of the cost function shrinks exponentially with the number of qubits.
Various issues lead to barren plateaus, such as deep ansatzes that lack structure~\cite{mcclean2018barren,sharma2020trainability,holmes2021connecting}, global cost functions~\cite{cerezo2020cost,sharma2020trainability}, high levels of noise~\cite{wang2020noise,du2020learnability}, scrambling target unitaries~\cite{holmes2020barren}, and large entanglement~\cite{marrero2020entanglement,patti2020entanglement}.

Without effort to avoid barren plateaus, this phenomenon can have a major impact on the scaling of one's algorithm. Specifically, the exponential suppression of the gradient implies that one would need an exponential precision to make progress in  %gradient descent
the optimization, consequently, causing one's algorithm to scale exponentially in the number of qubits. The standard goal of quantum algorithms is polynomial scaling, unlike the exponential scaling of classical algorithms. Hence, the exponential scaling due to barren plateaus could erase the possibility of a quantum speedup with a parametrized quantum circuit. It is therefore crucial to study barren plateaus in VQAs and QNNs in order to understand when quantum speedup is possible.

This has spawned an important research direction of finding strategies to avoid barren plateaus. Some examples include employing local cost functions~\cite{cerezo2020cost}, modifying the architecture~\cite{pesah2020absence,zhang2020toward}, pre-training~\cite{verdon2019learning}, parameter correlation~\cite{volkoff2021large}, layer-by-layer training~\cite{skolik2020layerwise}, and initializing layers to the identity~\cite{grant2019initialization}. These strategies are promising. However, more analytical and numerical studies are needed to understand how effective they are in general, for example, as in Ref.~\cite{campos2021abrupt}.

One possible strategy to consider is the choice of optimizer. It is widely believed that gradient-based optimizers will be directly impacted by barren plateaus, for obvious reasons. Moreover, higher-order derivatives are also exponentially suppressed in a barren plateau~\cite{cerezo2020impact}, so optimizers based on such derivatives will also be impacted. Nevertheless, there still remains the question of whether gradient-free optimizers could somehow avoid the barren plateau problem. This is currently a topic of debate~\cite{cerezo2020cost,marrero2020entanglement}. The question is naturally made subtle by the fact that gradient-free optimizers can potentially use global information about the landscape, rather than being restricted to using local gradient information.

In this work, we present an analytical argument suggesting that gradient-free approaches will, indeed, be impacted by barren plateaus.  Specifically, we show that cost function differences, $C(\vec{\theta}_B)-C(\vec{\theta}_A)$, will be exponentially suppressed in a barren plateau. This holds even when the points $\vec{\theta}_A$ and $\vec{\theta}_B$ are not necessarily close in parameter space. Gradient-free optimizers use such cost function differences to make decisions during the optimization. Hence, our results imply that such optimizers will either need to spend exponentially large resources to characterize cost function differences, or else these optimizers will not make progress in the optimization.

We confirm our analytical results with numerical simulations involving several gradient-free optimizers: Nelder-Mead, Powell, and COBYLA. For each of these optimizers, we attempt to train a deep parametrized quantum circuit, corresponding to the barren plateau scenario in Ref.~\cite{mcclean2018barren}. In all cases, we find that the number of shots (i.e., the amount of statistics) required to begin to train the cost function grows exponentially in the number of qubits. This is the same behavior that one sees for gradient-based methods, and is a hallmark of the barren plateau phenomenon.

\section{Theoretical Background}

Here we provide background needed to understand our results. We first consider the cost function used to train parameterized quantum circuits. Then we consider optimizers that can be used to optimize this cost function, with a specific focus on gradient-free optimizers. Finally, we give background on the barren plateau phenomenon.

\subsection{Cost function}

Consider a parameterized quantum circuit $V(\thv)$, whose parameters will be trained by minimizing a cost function $C(\thv)$. In this work, we consider a highly general cost function that can be expressed in the form
\begin{equation}\label{eq:cost}
    C(\thv) =\sum_{x=1}^{S} f_x(\thv,\rho_x)\,.
\end{equation}
Here, $\thv$ is a vector of $m$ continuous parameters, $\{\rho_x\}_{x=1}^S$ are $n$-qubit input quantum states from a training set $\SC$ of size $S$, and $f_x$ are functions that encode the problem and which can be different for each input state.

To ensure algorithmic efficiency, we assume that the number $m$ of parameters in $\thv$ is in $\OC(\poly(n))$. In addition we consider that any $\theta_\mu\in\thv$ parametrizes a  unitary of the form $e^{-i\theta_\mu H_{\mu}}$. We assume $H_{\mu}$ is a Hermitian operator with two distinct non-zero eigenvalues (e.g., $H_{\mu}$ could be a Pauli operator).

We remark that the cost function in~\eqref{eq:cost} contains as special cases many relevant applications. For instance, in a binary classification problem the cost function is given by the mean squared-error $C(\thv)=\sum_{x}(y_x-\widetilde{y}(\thv,\rho_x))^2$~\cite{farhi2018classification,killoran2019continuous,beer2020training,cong2019quantum}. Here, the training set is given by $\SC=\{\rho_x,y_x\}$ where $y_x$ are the true labels, and $\widetilde{y}(\thv,\rho_x)$ are the labels predicted by the Quantum Neural Network. In addition, the cost of several Variational Quantum  Algorithms is covered by~\eqref{eq:cost}. In this case, the cost takes a simpler form, where the training set contains a single state ($S=1$) and the cost is $C(\thv)=\Tr[OV(\thv)\rho V\ad(\thv)]$, with $O$ a Hermitian operator~\cite{peruzzo2014variational,mcclean2016theory,farhi2014quantum,romero2017quantum,khatri2019quantum,larose2019variational,arrasmith2019variational,cerezo2020variationalfidelity,cirstoiu2020variational,sharma2019noise,bravo2020variational,cerezo2020variational}.

The goal is then to solve the optimization problem
\begin{equation}\label{eq:optimization}
    \thv_{\opt}=\argmin_{\thv} C(\thv)\,.
\end{equation}
This involves choosing an optimizer, which can either be a gradient-based or gradient-free optimizer. Various gradient-based approaches~\cite{kubler2020adaptive,sweke2020stochastic,arrasmith2020operator,stokes2020quantum} have been proposed for training parameterized quantum circuits, and these will be directly impacted by barren plateaus.  Optimizers employing higher-order derivatives are also impacted by barren plateaus~\cite{cerezo2020impact}. In this work we consider the case when one employs a gradient-free optimization method. In the next section we review some widely-used gradient-free optimizers.

\subsection{Gradient-Free Optimizers}

We will refer to any optimization method that only accesses a zeroth-order oracle (i.e., does not directly access derivative information) as being gradient free. This is a very large class of methods, but they all depend on being able to distinguish cost function values at different points. Though our analytical results are general to any such optimizer, we now introduce three particular gradient-free optimizers that we will examine numerically: Nelder-Mead, Powell's Method, and COBYLA.

\subsubsection{Nelder-Mead}

\begin{figure*}[ht]
\includegraphics[width=1.9\columnwidth]{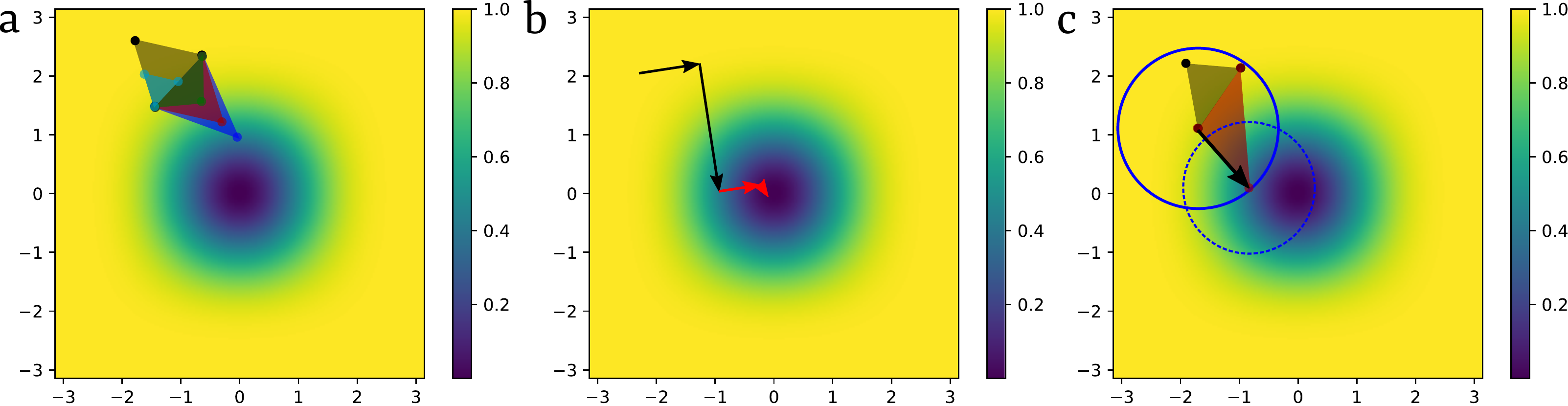}
\caption{Graphical depiction of the gradient-free optimizers considered. Panel \textbf{a} shows the different operations that the Nelder-Mead algorithm performs on the initial (grey) simplex: reflection (red), reflection with expansion (blue), reflection with contraction (green), and shrinking (turquoise). Panel \textbf{b} shows two iterations of the Powell method (with black for the first iteration and red for the second). Note that the direction of the final step is changed, reflecting a modified search direction. Finally, panel \textbf{c} shows an illustration of a COBYLA step from an initial (grey) simplex. After fitting a plane to the initial simplex, the method steps along the fitted slope to form a new simplex (red). The trust region is shown as a solid blue circle. A smaller trust region which might be used later in the optimization is illustrated with the dashed blue circle (though for this particular step the trust region would likely not be contracted).} 
\label{fig:opts}
\end{figure*}

One popular gradient-free optimization strategy is the Nelder-Mead algorithm~\cite{nelder1965simplex}. In this approach, one constructs a simplex in the space to be optimized over. Then one modifies it with a sequence of reflect, expand, contract, and shrink operations to move the simplex and then shrink it around the minimum. These operations are chosen based on conditional comparisons of the cost function values at each vertex as well as proposed new vertices. See Figure~\ref{fig:opts}\textbf{a} for an illustration these operations.

When used in an environment where the errors in those cost function values are large enough to cause mistakes in these comparisons, however,  this algorithm is vulnerable to performing shrink operations prematurely, which slows the optimization down and may lead to a false appearance of convergence~\cite{barton1991modifications}. Due to this difficulty, one would expect the number of iterations required to converge with Nelder-Mead to be especially bad in limited precision environments, though we note that there are a number of modifications that attempt to improve the method's robustness to noise~\cite{barton1991modifications,huang2018robust}. 

\subsubsection{Powell's Method}

The Powell algorithm~\cite{powell1964efficient} is another popular gradient-free optimizer that performs sequential line searches. This method starts with some input set of search vectors $V=\{\vec{v}_i\}$, usually just the coordinate directions in the parameter space. Searching along each of these directions in sequence, this method looks for the displacement $\{a_i\}$ along each direction that would minimize the cost when only varying parameters along the current direction. Finding the displacements $\{a_i\}$ is typically done with Brent's parabolic interpolation method~\cite{brent2013algorithms}, though in principle one could use any univariate gradient free optimizer. 

After sweeping through all of the search vectors, the iteration is completed by replacing the search vector $\vec{v}_j$ that corresponds to the greatest displacement, $a_j=\max({a_i})$, and replacing it with 
\begin{equation}
    \vec{v}_j\to \sum_i a_i \vec{v}_i.
\end{equation}
By making this replacement, convergence is accelerated and the method avoids getting stuck in a cyclic pattern of updates. See Figure~\ref{fig:opts}\textbf{b} for a sketch of two iterations of this method.

\subsubsection{COBYLA}

Constrained Optimization BY Linear Approximation (COBYLA) is another popular gradient-free optimizer by Powell~\cite{powell1994direct}. This algorithm constructs a simplex and uses the $m + 1$ points in the parameter space, with $m$ being the number of parameters, to define a hyperplane to capture the local slope of the cost function. The algorithm then replaces the highest cost function value point on the simplex by stepping from the lowest cost point along the direction of the slope. The method steps as far as possible along this estimated slope while staying within a lower bound on the radius of the trust region. 

The lower bound on the size of the trust region is decreased when the algorithm detects that is has stopped making progress, allowing the method to converge. Note, however, that the size of the trust region never increases in COBYLA. An iteration of this method (showing a shrinking trust region) is sketched in Figure~\ref{fig:opts}\textbf{c}.

\subsection{Barren Plateaus}

When the cost function exhibits a barren plateau, the cost function gradient vanishes exponentially with the system size. Without loss of generality we consider here the following generic definition of a barren plateau. 
\begin{definition}[Barren Plateau]\label{def:BP}
Consider the cost function defined in Eq.~\eqref{eq:cost}. This cost exhibits a barren plateau if, for all $\theta_\mu\in\thv$, the expectation value of the cost function partial derivative $\partial C(\thv)/\partial\theta_\mu=\partial_\mu C(\thv)$ is $\text{E}_{\thv}[\partial_\mu C(\thv)]=0$ and its variance vanishes exponentially with the number of qubits $n$ as
\begin{equation}\label{eq:var}
    \Var_{\thv}[\partial_\mu C(\thv)]\leq F(n)\,,\quad \text{with}\quad F(n)\in\OC\left(\frac{1}{b^n}\right)\,.
\end{equation}
for some $b> 1$. As indicated, the expectation values are taken over the parameters $\thv$.
\end{definition}

We remark here that, as shown in Definition~\ref{def:BP}, the barren plateau phenomenon is a probabilistic statement. In fact, from Chebyshev's inequality we know that $\Var_{\thv}[\partial_\mu C(\thv)]$  bounds the probability that the cost function partial derivative deviates from its mean of zero as
\begin{equation}\label{eq:cheb}
    P(|\partial_\mu C(\thv)|\geq c)\leq \frac{\Var_{\thv}[\partial_\mu C(\thv)]}{c^2}\,,
\end{equation}
for any $c>0$. In practice this means that by randomly initializing the parameters $\thv$, there is a high probability that one ends up in a flat region of the landscape where the gradients are exponentially suppressed.

Let us now discuss different mechanisms that can lead to barren plateaus in the cost function landscape. As shown in the seminal work of Ref.~\cite{mcclean2018barren}, deep random unstructured circuits which form $2$-designs on $n$ qubits will exhibit barren plateaus. Here we use the term deep when the depth of the ansatz is in $\OC(\poly(n))$. For instance, as shown in~\cite{harrow2009random,brandao2016local,harrow2018approximate} local circuits will form $2$-designs when their depth is in $\OC(\poly(n))$.

The barren plateau phenomenon was extended in~\cite{cerezo2020cost} to a type of shallow depth ansatz known as the layered hardware efficient ansatz, where random local gates act on alternating pairs of neighboring qubits in a brick-like structure.  Here it was shown that the locality of the cost function can be linked to its trainability. Specifically, global cost functions (those where one compares operators living in exponentially large Hilbert spaces) exhibit barren plateaus for any circuit depth. On the other hand, it was shown that local cost functions (where one compares operators on an individual qubit level) are trainable when the ansatz depth is in $\OC(\log(n))$, as here their gradients vanish at worst polynomially  (rather than exponentially) with the system size. 

Barren plateaus have also been shown to arise in more general QNN architectures~\cite{sharma2020trainability,marrero2020entanglement}. In perceptron-based QNNs with hidden and visible layers, connecting a large number of qubits in different layers with random global perceptrons (and hence highly entangling them) can lead to exponentially vanishing gradients. These results have shown that the barren plateau phenomenon is a generic problem that can arise in multiple architectures for quantum machine learning. 

Finally, in~\cite{wang2020noise} a noise-induced barren plateau  mechanism was found. Here it was proven that the presence of noise acting before and after each unitary layer in a parametrized quantum circuit leads to exponentially vanishing gradients for circuits with linear or super-linear depth. When the cost exhibits a noise-induced barren plateau we have $|\partial_\mu C(\thv)|\leq \widehat{F}(n)$ with $\widehat{F}(n)\in\OC(1/\widehat{b}^n)$ for some $\widehat{b}>1$. The underlying mechanism here is that the state gets corrupted due to noise, leading to a flattening of the whole cost landscape. This phenomenon is conceptually different from the previous barren plateaus as here one does not average over the parameters $\thv$. Nevertheless, the noise-induced barren plateau still satisfies Definition~\ref{def:BP}, which is a weaker condition.

\section{Main Results}

In this section we first present our main analytical results in the form of Proposition~\ref{prop:1} and Corollary~\ref{cor:1}. We then discuss the implications for employing gradient-free optimizers in a barren plateau. 

\subsection{Exponentially suppressed cost differences}

Here we consider two relevant scenarios where we analyze, on average, how large the difference $\Delta C= C(\thv_B)-C(\thv_A)$ between two points in the landscape can be. First we consider the case when  $\thv_A$ and $\thv_B$ are not independent, but rather $\thv_B$ can be obtained from $\thv_A$ through a given translation in parameter space. We then  analyze the case when $\thv_A$ and $\thv_B$ are independent.

The following proposition constitutes the main result of our work. The proof is presented in the Appendix. 

\begin{proposition}\label{prop:1}
Consider the cost function of Eq.~\eqref{eq:cost}. Let $\thv_A$ be a randomly chosen point in parameter space. Let  $\thv_B=\vec{\theta}_A+L\hat{\vec{\ell}}$ be  a point at a distance  $L=\|\vec{\theta}_B-\vec{\theta}_A\|$ from $\thv_A$ in parameter space, for some unit vector $\hat{\vec{\ell}}$. If the cost exhibits a barren plateau according to Definition~\ref{def:BP}, then the expectation value of the difference $\Delta C=C(\thv_B)-C(\thv_A)$ is
\begin{equation}
    \text{E}_{\vec{\theta}_A}[\Delta C]=0\,,
\end{equation}
and the variance is exponentially vanishing with $n$ as
\begin{equation}
    \Var_{\vec{\theta}_A}[\Delta C]\leq G(n)\,,
\end{equation}
with 
\begin{equation}\label{eq:G-n}
    G(n)=m^2L^2 F(n)\,, \quad \text{and}\quad G(n)\in \widetilde{\OC}\left(\frac{1}{b^n}\right)\,,
\end{equation}
for some $b> 1$. Here $m$ is the dimension of the parameter space,  and $F(n)$ was defined in~\eqref{eq:var}.
\end{proposition}

Let us here recall that  we have assumed that $m\in\OC(\poly(n))$. Similarly, we have that $\theta_\mu$ parametrizes a unitary generated by a Hermitian operator $H_{\mu}$ with two distinct non-zero eigenvalues. From the latter it then follows that $L$ is always in $\OC(\poly(n))$, and hence that $G(n)\in \widetilde{\OC}(1/b^n)$. 

From the previous results one can readily evaluate the case when $\thv_B$ and $\thv_A$ are independent. This case is of relevance to global optimizers, such as Bayesian approaches, where initial points on the landscape are chosen independently. This scenario can be analyzed by computing the expectation value $\text{E}_{\vec{\theta}_A,\vec{\theta}_B}[\Delta C]=\text{E}_{\vec{\theta}_B}[\text{E}_{\vec{\theta}_A}[\Delta C]]$. From Proposition~\ref{prop:1}, we can derive the following corollary.

\begin{corollary}\label{cor:1}
Consider the cost function of Eq.~\eqref{eq:cost}. Let $\thv_A$ and $\thv_B$ be two randomly chosen points in parameter space. Without loss of generality we assume that $\thv_B=\vec{\theta}_A+L\hat{\vec{\ell}}$ for random $L$ and $\hat{\vec{\ell}}$ so that $\text{E}_{\vec{\theta}_A,\vec{\theta}_B}[\cdots]=\text{E}_{\vec{\theta}_A,L,\hat{\vec{\ell}}}[\cdots]$. If the cost exhibits a barren plateau according to Definition~\ref{def:BP}, then the expectation value of the difference $\Delta C=C(\thv_B)-C(\thv_A)$ is $\text{E}_{\vec{\theta}_A,L,\hat{\vec{\ell}}}[\Delta C]=0$, 
and the variance is exponentially vanishing with $n$ as
\begin{equation}\label{eq:boundprop}
    \Var_{\vec{\theta}_A,L,\hat{\vec{\ell}}}[\Delta C]\leq \widehat{G}(n)\,,
\end{equation}
with 
\begin{equation}
   \widehat{G}(n)=m^2\overline{L}^2 F(n)\,, \quad \text{and}\quad \widehat{G}(n)\in \widetilde{\OC}\left(\frac{1}{b^n}\right)\,,
\end{equation}
for some $b> 1$. Here $m$ is the dimension of the parameter space,  $F(n)$ was defined in~\eqref{eq:var}, and
\begin{equation}
    \overline{L}=\text{E}_{L,\hat{\vec{\ell}}}\left[L\right]
\end{equation}
is the average distance between any two points in parameter space.
\end{corollary}

The proof of Corollary~\ref{cor:1} readily follows from Proposition~\ref{prop:1} by additionally computing the expectation value over $L$ and $\hat{\vec{\ell}}$.  Moreover, here we can see that $\widehat{G}(n)$ is exponentially vanishing with the system size as  $\overline{L}\in\OC(\poly(n))$. 

From Proposition~\ref{prop:1} we have that given two dependent set of parameters $\thv_A$ and a set $\thv_B$ related trough a translation in parameter space, then the probability that the difference $\Delta C=C(\thv_B)-C(\thv_A)$ is larger than a given $c>0$ can be bounded as 
\begin{equation}\label{eq:difference-bound}
    P(|\Delta C|\geq c)\leq \frac{G(n)}{c^2}\,,
\end{equation}
where we have used~\eqref{eq:cheb}, and Eq.~\eqref{eq:boundprop} from Proposition~\ref{prop:1}. Note that a similar result can be obtained for the case when $\thv_A$ and $\thv_B$ are independent, but here one replaces $G(n)$ by $\widehat{G}(n)$ in~\eqref{eq:difference-bound}.  This implies that, with high probability, the difference $\Delta C$   will be exponentially vanishing with the system size, for both cases when $\thv_A$ and $\thv_B$ are dependent or independent. Moreover, we remark that this is a direct consequence of the fact that the cost exhibits a barren plateau.

\subsection{Implications for gradient-free optimizers}

Let us first recall that, as discussed in the previous section, the capability of distinguishing the cost function value at different sets of parameters is at the core of gradient-free optimization methods. Therefore, the precision required to differentiate choices fundamentally limits the scaling of these methods, with smaller differences requiring greater precision. If an optimizer's precision requirements are not met, then each decision the method makes becomes randomly chosen by shot noise, leading to many optimizers effectively becoming either random walks or random sampling.

The results in Proposition~\ref{prop:1} pertain to gradient-free optimizers that compare points that are a given distance and direction apart. For example, simplex-based methods like Nelder-Mead fall under this category. As we show that cost differences are exponentially suppressed with the system size in a barren plateau, this leads to applications having sampling requirements that scale exponentially. Exponentially scaling sampling requirements, in turn, hinder the possibility of achieving quantum speedup with such an algorithm.

Similarly, Corollary~\ref{cor:1} tells us that cost function differences between randomly chosen points are also exponentially suppressed. This means that either random search methods or methods that use random initialization, such as Bayesian optimization~\cite{movckus1975bayesian}, will also struggle with barren plateau landscapes. Therefore, using randomness in the selection of points cannot evade this exponential scaling result.

Let us finally remark that Proposition~\ref{prop:1} and Corollary~\ref{cor:1} make no assumption about how close (or far) the parameters $\thv_A$ and $\thv_B$ are in parameter space other than that $L=\|\vec{\theta}_B-\vec{\theta}_A\|\in\OC(\poly(n))$. Given that for any practical application the number of parameters $m$ should scale no faster than $m\in\OC(\poly(n))$ (or the problem will become untrainable for reasons having nothing to do with barren plateaus), this seems very reasonable. For example, if all of the parameters are single qubit rotations, the parameter space is a $m$-dimensional torus with unit radius. On that torus, the greatest length of the shortest path between two points is:
\begin{equation}
\begin{aligned}
    L_{\mathrm{max}}=&\max_{\thv_A,\thv_B}\|\thv_A-\thv_B\|\\
    =&\sqrt{m}\pi.
\end{aligned}
\end{equation}
 This means that our results are valid for both local and global optimizers as sampling points that are further apart cannot overcome the suppressed slope.

\section{Numerical Implementation}

\begin{figure}[t]
\includegraphics[width=\columnwidth]{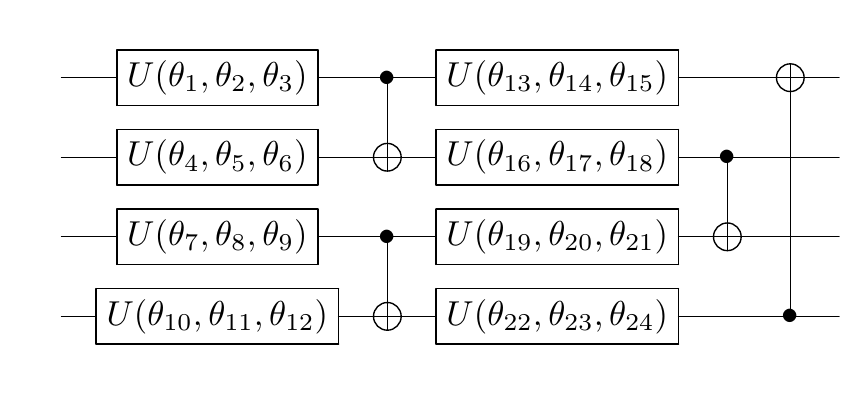}
\caption{Single layer of the hardware efficient ansatz employed in our numerical implementations, shown here for $n=4$. The $U$ gates  are general single qubit unitaries   $U(\theta_1,\theta_2,\theta_3) = R_Z(\theta_2+\pi)R_X(\pi/2)R_Z(\theta_1+\pi)R_X(\pi/2)R_Z(\theta_3)$. Here $R_Z(\alpha) = e^{-i\alpha/2 \sigma_Z}$,  $R_X(\alpha) = e^{-i\alpha/2 \sigma_X}$, and $\sigma_Z$, $\sigma_X$ are Pauli matrices.     } 
\label{fig:hardeff}
\end{figure}

In this section we present numerical results obtained by simulating a variational quantum compiling algorithm~\cite{khatri2019quantum,sharma2019noise}. Here, one trains a parametrized quantum circuit $V(\thv)$ to approximate a target unitary $U$. Specifically, we consider the toy-model problem where $U=\id$ and the goal is to train the parameters in $V(\thv)$ such that $V(\thv)\ket{\vec{0}}=\ket{\vec{0}}$, with $\ket{\vec{0}}=\ket{0}^{\otimes n}$ the all-zero state. As shown in~\cite{khatri2019quantum,sharma2019noise}, the following local cost function is faithful
\begin{equation}
C(\thv) = \Tr[ O_L V(\thv)\dya{\vec{0}}V\ad(\thv)]\,,
\label{cost_function}
\end{equation}
where
\begin{equation}
    O_L = \id - \frac{1}{n}\sum_{i=1}^{n}\ket{0_i}\bra{0_i}\,,
\end{equation}
in that can verify that $C(\thv)\in[0,1]$, with $C(\thv)=0$ iff $V(\thv)\ket{\vec{0}}=\ket{\vec{0}}$ (up to global phase). We remark that there is an efficient quantum circuit to compute $C(\thv)$~\cite{sharma2019noise}.

For $V(\thv)$ we employ a layered hardware efficient ansatz as shown in Fig.~\ref{fig:hardeff}. Moreover, we recall that Ref.~\cite{mcclean2018barren} showed that a cost function such as~\eqref{cost_function} with a layered hardware efficient ansatz that is randomly initialized will exhibit barren plateaus when the depth of $V(\thv)$ scales at least linearly in $n$.

In our numerics we simulated the aforementioned quantum compilation  task for different numbers of qubits $n=5,6,\ldots,11$. Letting $p$ be the number of layers in the ansatz in Fig.~\ref{fig:hardeff}, we choose $p=n$, so that the depth grows linearly in $n$. This corresponds to the barren plateau scenario in Ref.~\cite{mcclean2018barren}. For each value of $n$, we solved the optimization of Eq.~\eqref{eq:optimization} by employing the Nelder-Mead, Powell, and COBYLA methods. These simulations were performed using MATLAB (Nelder-Mead) and SciPy (Powell and COBYLA). In all cases we randomly initialized the parameters $\thv$ in the ansatz and we ran the optimizer until a cost function value $C=0.4$ was achieved or until a maximal total number of shots used throughout the optimization was surpassed. For simplicity we use the default values for  hyper-parameters  not related to optimization termination.   We note that we chose a relatively large value ($C=0.4$) for the cost threshold because we are interested in the initial stages of the training process, i.e., the question of whether one can get past the barren plateau. With this choice, the computational expense of reaching this threshold does not take into account the difficulty of finding a minimum, it only reflects the burden of the barren plateau.

Since the goal is to heuristically determine the precision (i.e., the number of shots $N$) needed to minimize the cost, we first ran simulations with different values of  $N$ allocated per cost-function evaluation. For each $N$ we simulated $20$ optimization instances (runs) with different initial points and we kept track of  $N_{\text{total}}$ used throughout the optimization. The next step was to determine the value of $N$ for which a cost of $C=0.4$ could be reached and which minimizes the median total number of shots $N_{\text{total}}$  computed over different runs. We analyze the scaling of this median value as a function of $n$ below.   

 \begin{figure}[t]
\includegraphics[width=\columnwidth]{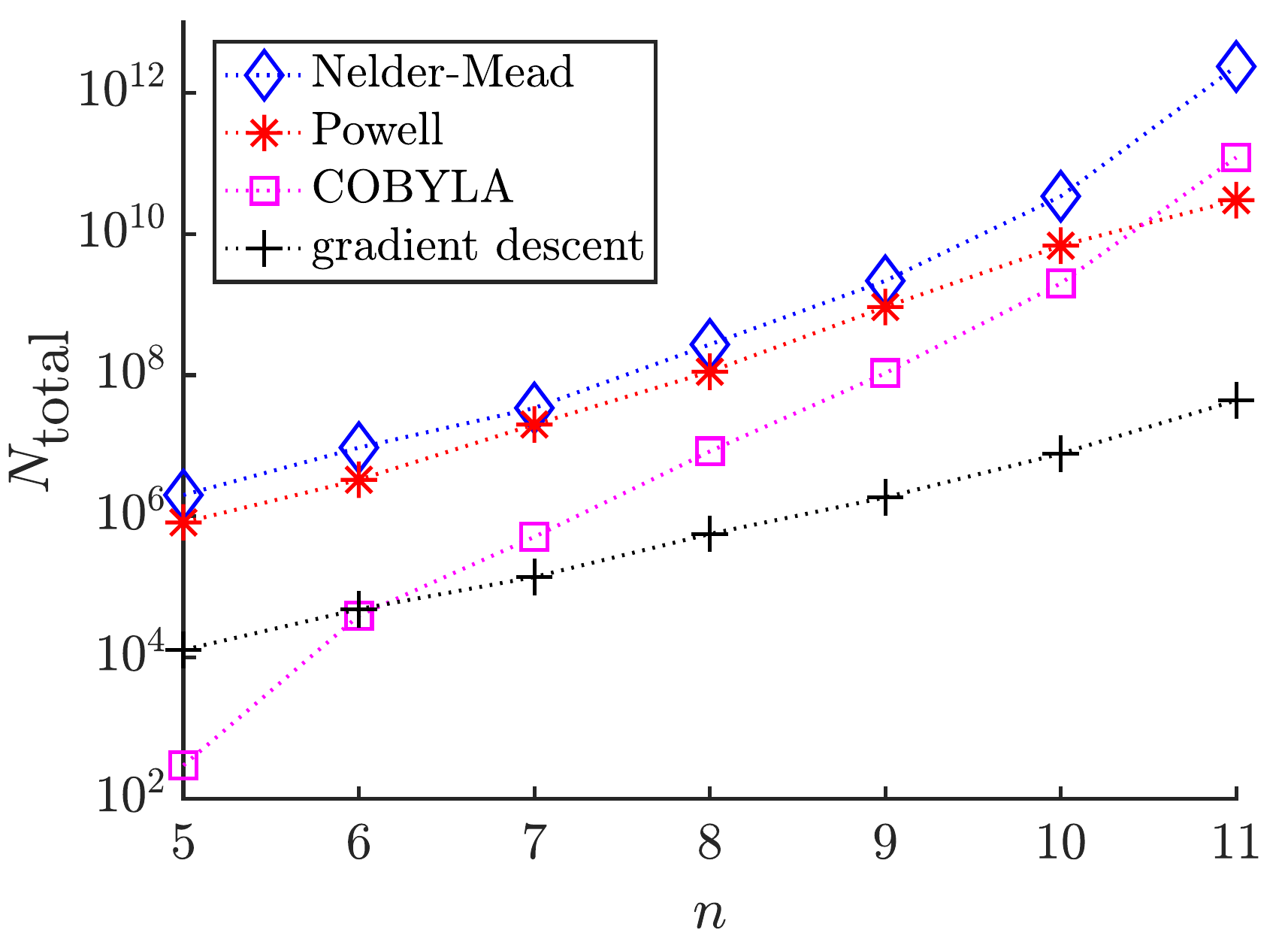}
\caption{Median value of the total shot number $N_{\textrm{total}}$ necessary to reach $C=0.4$ plotted versus the number of qubits $n$. We show results for  Nelder-Mead  (blue diamonds), Powell (red asterisks), and COBYLA (pink squares) implementations, as well as results for a gradient-descent implementation (black crosses) for reference. Each data point was obtained from a sample of $20$ runs initialized by random initial angles~$\vec{\theta}$.
}
\label{fig:numerical}
\end{figure}

In Fig.~\ref{fig:numerical} we present our numerical results.  Here we can see that the total number of shots scales exponentially with $n$ for the Powell method, and scales super-exponentially for the Nelder-Mead optimizer. For the COBYLA method, the  $N_{\text{total}}$ behavior as a function of $n$ is not very regular but it is consistent with at least   an exponential increase. As a reference point we also show in Fig. \ref{fig:numerical}  results obtained with a custom gradient-descent optimizer. As expected, the total number of shots also scales exponentially in this case.

\section{Discussion}

With a wide range of applications spanning chemistry, optimization, and big data analysis, training parameterized quantum circuits is arguably the leading paradigm for near-term quantum computing. Yet barren plateaus in the training landscape remains an obstacle to making these paradigms scalable. Hence, one of the most important lines of research in this field is developing methods to avoid barren plateaus.

In this work, we consider the question of whether the choice of optimizer could be a potential strategy in avoiding barren plateaus. We focus on gradient-free optimizers, since there has been recent debate in the community about whether barren plateaus effect such optimizers.

Our main result is an analytical argument suggesting that gradient-free optimizers will, indeed, be impacted by barren plateaus. Proposition~\ref{prop:1} is relevant to gradient-free optimizers that search through the landscape starting from a (random) initial point. For example, this includes simplex-based optimizers like Nelder-Mead. This proposition asserts that the variance of cost function differences is exponentially suppressed in a barren plateau. This implies that such optimizers will need to expend exponentially large resources in order to make decisions about where to move in the landscape.

Corollary~\ref{cor:1} considers a slightly different scenario, where both points are randomly and independently chosen. This is relevant to global gradient-free optimizers, such as Bayesian methods, which initially choose multiple random points on the landscape, and then proceed from this initial set of points. This corollary implies that these optimizers will also need to utilize exponentially large resources in order to make progress.

We also numerically attempt to train in a barren plateau scenario using several gradient-free optimizers. We ask how many shots are required to begin to train the cost function. In all cases, we find that the required number of shots grows exponentially in the number of qubits. This is consistent with our main result, and demonstrates that barren plateaus can lead to exponential scaling even for gradient-free optimizers. We note that this exponential scaling is a lower bound on the asymptotics. For the Nelder-Mead we find super-exponential scaling, likely due to the chances of prematurely shrinking the simplex when it is hard to order the cost values~\cite{barton1991modifications}. For the case of COBYLA, there may be a similar effect from prematurely shrinking the radius of the trust region, though this is not clearly demonstrated in our data. Finally, the Powell method appears to show exponential scaling. It is likely that the reason Powell shows better scaling is that, unlike the other optimizers, statistical noise does not have a cumulative effect on the state of the optimizer.

Our work casts doubt on the notion that the choice of optimizer could provide a strategy to avoid barren plateaus. While the asymptotically exponential scaling cannot be avoided, we note that the size limits of trainable problems may be extended by a careful choice of optimization strategy. For example, techniques using neural networks~\cite{verdon2019learning} or natural evolutionary strategies~\cite{anand2021natural} may improve the constants multiplying the exponential scaling. However, we emphasize that all such strategies at minimum require the comparison of cost function values at different points and thus are subject to our scaling analysis. 

This result highlights the difficult challenge posed by barren plateaus. Future work certainly should continue to develop strategies to avoid them. Additionally, in future work, we hope to develop a unified treatment that covers the impact of barren plateaus on various types of optimizers, gradient-based and gradient-free.

%    |\---/|
%    | ,_, |
%     \_`_/-..----.
%  ___/ `   ' ,""+ \  
% (__...'   __\    |`.___.';
%   (_,...'(_,.`__)/'.....+

\section*{Acknowledgements}

AA and LC were initially supported by LDRD program of LANL under project number 20190065DR. MC acknowledges support from the Center for Nonlinear Studies at Los Alamos National Laboratory (LANL). Piotr C. was supported by the Laboratory Directed Research and Development (LDRD) program of Los Alamos National Laboratory (LANL) under project number 20190659PRD4.  PJC acknowledges initial support from the LANL ASC Beyond Moore's Law project. This work was supported by the U.S. DOE, Office of Science, Office of Advanced Scientific Computing Research, under the Accelerated Research in Quantum Computing (ARQC) program.

\bibliographystyle{apsrev4-1mod}
\bibliography{quantum.bib}

\newpage
\pagebreak
\onecolumngrid

\setcounter{section}{0}
\setcounter{proposition}{0}
\setcounter{theorem}{0}
\setcounter{corollary}{0}

\section*{Appendix}

\appendix

\label{app:grad_impl}

\section{Proof of Proposition~\ref{prop:1}}
We now prove our main result. We restate the proposition here for convenience. 

\begin{proposition}
Consider the cost function of Eq.~\eqref{eq:cost}. Let $\thv_A$ be a randomly chosen point in parameter space. Let  $\thv_B=\vec{\theta}_A+L\hat{\vec{\ell}}$ be  a point at a distance  $L=\|\vec{\theta}_B-\vec{\theta}_A\|$ from $\thv_A$ in parameter space, for some unit vector $\hat{\vec{\ell}}$. If the cost exhibits a barren plateau according to Definition~\ref{def:BP}, then the expectation value of the difference $\Delta C=C(\thv_B)-C(\thv_A)$ is
\begin{equation}
    \text{E}_{\vec{\theta}_A}[\Delta C]=0\,,
\end{equation}
and the variance is exponentially vanishing with $n$ as
\begin{equation}
    \Var_{\vec{\theta}_A}[\Delta C]\leq G(n)\,,
\end{equation}
with 
\begin{equation}
    G(n)=m^2L^2 F(n)\,, \quad \text{and}\quad G(n)\in \widetilde{\OC}\left(\frac{1}{b^n}\right)\,,
\end{equation}
for some $b> 1$. Here $m$ is the dimension of the parameter space,  and $F(n)$ was defined in~\eqref{eq:var}.
\end{proposition}

\textit{Proof.} We begin by noting that finite differences in cost values can be expressed as:
\begin{equation}
\begin{aligned} \Delta C=&C(\vec{\theta}_B)-C(\vec{\theta}_A)\\
=&\int_{\vec{\theta}_A}^{\vec{\theta}_B}\vec{\nabla}C(\vec{\theta})\cdot d\vec{\theta}\,.
\end{aligned}
\end{equation}

Note that as we are integrating over gradients (which form a conservative vector field by definition) this integral is path independent. We can therefore choose to integrate along the line segment between $\vec{\theta}_B$ and $\vec{\theta}_A$, setting $\vec{\theta}=\vec{\theta}_A+\ell\hat{\vec{\ell}}$, with $\hat{\vec{\ell}}$ being the unit vector along $\vec{\theta}_B-\vec{\theta}_A$ and $\ell \in [0,L]$ where $L=\|\vec{\theta}_B-\vec{\theta}_A\|$. We then have $d\vec{\theta}=\hat{\vec{\ell}}d\ell$. To show that the mean difference is zero, we simply take the expectation value of this integral with respect to $\thv_A$
\begin{equation}
\begin{aligned} 
\text{E}_{\vec{\theta}_A}[\Delta C]=&\int_{0}^{L}\text{E}_{\vec{\theta}_A}[\vec{\nabla}C(\vec{\theta}_A+\ell\hat{\vec{\ell}})\cdot\hat{\vec{\ell}}] d\ell\\
=&0\,.
\end{aligned}
\end{equation}
This follows as averaging $\thv_A$ over parameter space is equivalent to averaging $\thv_A+\ell\hat{\vec{\ell}}$ over the same space (for any fixed $\ell\hat{\vec{\ell}}$), and the average gradient over the parameter space is zero from the definition of a barren plateau (see Definition~\ref{def:BP}).

Similarly, we can compute the magnitude of the second moment of this difference as
\begin{equation}
\begin{aligned} 
\big| \text{E}_{\vec{\theta}_A}[(\Delta C)^2]\big|
=&\Bigg|\int_{0}^{L}\int_{0}^{L}\text{E}_{\vec{\theta}_A}[(\vec{\nabla}C(\vec{\theta}_A+\ell\hat{\vec{\ell}})\cdot\hat{\vec{\ell}})(\vec{\nabla}C(\vec{\theta}_A+\ell'\hat{\vec{\ell}})\cdot\hat{\vec{\ell}})] d\ell d\ell'\Bigg|\\
=&\Bigg|\int_{0}^{L}\int_{0}^{L}\text{Cov}_{\vec{\theta}_A}\left((\vec{\nabla}C(\vec{\theta}_A+\ell\hat{\vec{\ell}})\cdot\hat{\vec{\ell}}),(\vec{\nabla}C(\vec{\theta}_A+\ell'\hat{\vec{\ell}})\cdot\hat{\vec{\ell}})\right) d\ell d\ell'\Bigg|\,.
\end{aligned}
\end{equation}
Note that we have dropped the square of the mean as the mean is zero, so the second moment is the covariance. Next, we upper bound this covariance using the product of the variances using the Cauchy-Schwartz inequality
\begin{equation}
\begin{aligned} 
\big| \text{E}_{\vec{\theta}_A}[(\Delta C)^2]\big|\le&\int_{0}^{L}\int_{0}^{L}\left(\text{Var}_{\vec{\theta}_A}(\vec{\nabla}C(\vec{\theta}_A+\ell\hat{\vec{\ell}})\cdot\hat{\vec{\ell}})\text{Var}_{\vec{\theta}_A}(\vec{\nabla}C(\vec{\theta}_A+\ell'\hat{\vec{\ell}})\cdot\hat{\vec{\ell}})\right)^\frac{1}{2} d\ell d\ell'\,.
\end{aligned}
\end{equation}
Next, we split these variances into the covariances between individual components of the gradient
\begin{equation}
\begin{aligned} 
\big| \text{E}_{\vec{\theta}_A}[(\Delta C)^2]\big|\le&\int_{0}^{L}\int_{0}^{L}\left(\sum_{i=1}^m\sum_{j=1}^m\sum_{p=1}^m\sum_{q=1}^m\text{Cov}_{\vec{\theta}_A}\left(\partial_i C(\vec{\theta}_A+\ell\hat{\vec{\ell}})\hat{\ell}_i,\partial_j C(\vec{\theta}_A+\ell\hat{\vec{\ell}})\hat{\ell}_j\right)\right. \\
&\,\,\,\,\,\,\left.\text{Cov}_{\vec{\theta}_A}(\partial_p C(\vec{\theta}_A+\ell\hat{\vec{\ell}})\hat{\ell}_p,\partial_q C(\vec{\theta}_A+\ell\hat{\vec{\ell}})\hat{\ell}_q)\right)^\frac{1}{2} d\ell d\ell'\,.
\end{aligned}
\end{equation}
As before, these covariances can be bounded by variances using the Cauchy-Schwartz inequality
\begin{equation}
\begin{aligned} 
\big| \text{E}_{\vec{\theta}_A}[(\Delta C)^2]\big|\le&\int_{0}^{L}\int_{0}^{L}\Bigg(\sum_{i=1}^m\sum_{j=1}^m\sum_{p=1}^m\sum_{q=1}^m\left(\text{Var}_{\vec{\theta}_A}\left(\partial_i C(\vec{\theta}_A+\ell\hat{\vec{\ell}})\right)\text{Var}_{\vec{\theta}_A}\left(\partial_j C(\vec{\theta}_A+\ell\hat{\vec{\ell}})\right)\right)^\frac{1}{2}\\
&\,\,\,\,\,\,\left(\text{Var}_{\vec{\theta}_A}\left(\partial_p C(\vec{\theta}_A+\ell\hat{\vec{\ell}})\right)
\text{Var}_{\vec{\theta}_A}\left(\partial_q C(\vec{\theta}_A+\ell\hat{\vec{\ell}})\right)\right)^\frac{1}{2}\Bigg)^\frac{1}{2} d\ell d\ell'\,.
\end{aligned}
\end{equation}
If each variance can be bounded by a function $F(n)$, we can then write
\begin{equation}
\begin{aligned} 
\big| \text{E}_{\vec{\theta}_A}[(\Delta C)^2]\big|
\le&m^2\int_{0}^{L}\int_{0}^{L}F(n) d\ell d\ell'\\
=&m^2L^2F(n)\\
=&G(n).
\end{aligned}
\end{equation}

If the cost exhibits a barren plateau, then by Definition~\ref{def:BP},  $F(n)\in\OC\left(\frac{1}{b^n}\right)$ for some $b>1$. It then follows that if $m\in\OC(\poly(n))$ and $L\in\OC(\poly(n))$, then    $G(n)\in\OC\left(\frac{1}{b^n}\right)$ and Proposition~\ref{prop:1} holds.
\end{document}